\documentclass[aps,prd,twocolumns,eqsecnum,preprintnumbers,showpacs,amsmath,amssymb]
{revtex4}

\usepackage{graphicx}
\usepackage{dcolumn}
\usepackage{bm}

\begin{document}

\title{ENERGY FROM THE NONPERTURBATIVE QCD VACUUM}

\author{V. Gogokhia}
\email[]{gogohia@rmki.kfki.hu}

\affiliation{HAS, CRIP, RMKI, Depart. Theor. Phys., Budapest 114,
P.O.B. 49, H-1525, Hungary}

\date{\today}
\begin{abstract}
Using the effective potential approach for composite operators, we
have formulated a general method of calculation of the truly
nonperturbative Yang-Mills vacuum energy density (the Bag constant
apart from the sign, by definition). It is the main dynamical
characteristic of the QCD ground state. We define it as an
integration of the truly nonperturbative effective charge over the
nonperturbative region (soft momentum region). It is free of all
types of the perturbative contributions, by construction. For the
considered truly nonperturbative effective charge it is finite,
negative and it has no imaginary part (stable vacuum), as well as
it is a manifestly gauge-invariant, i.e., not explicitly depending
on the gauge-fixing parameter. A nontrivial minimization procedure
makes it possible to determine the Bag constant as a function
either of the mass gap, which is responsible for the large-scale
structure of the true QCD vacuum, or of the effective scale, which
separates the nonperturbative region from the perturbative one. We
have also argued that the Bag constant is a quant of the energy
density which can be released from the QCD vacuum, which, in its
turn, is considering as an infinite and permanent reservoir of
energy.
\end{abstract}



\maketitle

\section{Introduction}

The Lamb shift and the Casimir effect are probably the two most
famous experimental evidences of zero-point energy fluctuations in
the vacuum of Quantum Electrodynamics (QED) \cite{1,2}. The both
effects are rather weak, since the QED vacuum is mainly
perturbative (PT) by origin, character and magnitude (the
corresponding fine structure constant is weak). However, even in
this case there have been already made attempts to exploit the
Casimir effect in order to release energy from the vacuum (see,
for example Refs. \cite{3,4} and references in the above-mentioned
reviews \cite{1,2}). Let us also note that in Ref. \cite{5} by
investigating the thermodynamical properties of the quantum vacuum
it has been concluded that no energy can be extracted cyclically
from the vacuum (see, however Ref. \cite{2} and references
therein).

Contrary to the QED vacuum, the vacuum of Quantum Chromodynamics
(QCD) is a very complicated confining medium and its dynamical and
topological complexity \cite{6,7} means that its structure can be
organized at various levels: classical \cite{6,7,8} and quantum
\cite{9}. It is mainly nonperturbative (NP) by origin, character
and magnitude, since the corresponding fine structure constant is
strong. So the idea to exploit the NP QCD vacuum in order to
extract energy from it seems to be more attractive. However,
before to discuss the ways how to extract, it is necessary to
discuss which minimum/maxmimum amount of energy at all can be
released in a single cycle. The vacuum of QCD due to its
above-mentioned dynamical and topological complexity contains many
different components and ingredients. They contribute to the truly
NP vacuum energy density (VED), one of the main characteristics of
the QCD ground state. It is well known that the VED in general
badly diverges in quantum field theory, in particularly QCD. Thus
the main problem is how to correctly define the truly NP VED which
should be finite, negative and it should have no imaginary part
(stable vacuum). Being one of the main physical characteristics of
the QCD vacuum, the truly NP VED should not depend explicitly on
the gauge-fixing parameter as well.

Let us also remind that in order to calculate physical observables
from first principles in QCD it is not enough to know its
Lagrangian. It is also necessary and important to know the true
structure of its ground state. It is just the response of the true
QCD vacuum which substantially modifies all the QCD Green's
functions from their free counterparts. Just these full
("dressed") Green's functions are needed for the above-mentioned
calculations.

The main purpose of this paper is twofold. First of all, to define
correctly the truly NP VED, that is free of all types of the PT
contributions ("PT contaminations"), as well as to make it a
manifestly gauge-invariant quantity (not explicitly depending on
the gauge-fixing parameter). Secondly, to discuss in general terms
which amount of energy can be extracted from the QCD ground state,
which in its turn is considering as an infinite and permanent
reservoir of energy.

\section{The VED }

The quantum part of the VED is determined by the effective
potential approach for composite operators \cite{10,11}. In the
absence of external sources the effective potential is nothing but
the VED. It is given in the form of the loops expansion, where the
number of the vacuum loops (consisting in general of the confining
quarks and gluons properly regularized with the help of ghosts) is
equal to the power of the Plank constant, $\hbar$.

Here we are going to formulate a general method of numerical
calculation of the quantum part of the truly NP Yang-Mills (YM)
VED in the covariant gauge QCD. The gluon part of the VED which
to-leading order (log-loop level $\sim \hbar$) is given by the
effective potential for composite operators as follows \cite{10}:

\begin{equation}
V(D) =  { i \over 2} \int {d^4q \over {(2\pi)^4}}
 Tr\{ \ln (D_0^{-1}D) - (D_0^{-1}D) + 1 \},
\end{equation}
where $D(q)$ is the full gluon propagator and $D_0(q)$ is its free
counterpart (see below). The traces over space-time and color
group indices are understood. Evidently, the effective potential
is normalized to $V(D_0) = 0$ as usual. Next-to-leading and higher
contributions (two and more vacuum loops) are suppressed by one
order of magnitude in powers of $\hbar$ at least, and thus are not
important from the numerical point of view.

The two-point Green's function, describing the full gluon
propagator, is

\begin{equation}
D_{\mu\nu}(q) = - \left\{ T_{\mu\nu}(q)d(-q^2, \xi) + \xi
L_{\mu\nu}(q) \right\} {1 \over q^2 },
\end{equation}
where $\xi$ is the gauge-fixing parameter and $T_{\mu\nu}(q) =
g_{\mu\nu} - q_\mu q_\nu / q^2 = g_{\mu\nu } - L_{\mu\nu}(q)$. Its
free counterpart $D_0 \equiv D^0_{\mu\nu}(q)$ is obtained by
putting the full gluon form factor $d(-q^2, \xi)$ in Eq. (2.2)
simply to one, i.e., $D^0_{\mu\nu}(q) = - \left\{ T_{\mu\nu}(q) +
\xi L_{\mu\nu}(q) \right\} (1 / q^2)$. In order to evaluate the
effective potential (2.1), on account of Eq. (2.2), we use the
well-known expression, namely

\begin{equation}
Tr \ln (D_0^{-1}D) = 8 \times 4 \ln det (D_0^{-1}D) = 32 \ln [ (3/
4 )d(-q^2, \xi) + (1 / 4 ) ].
\end{equation}
It becomes zero (in accordance with the above-mentioned
normalization condition) when the full gluon form factor is
replaced by its free counterpart.

 Going over to four-dimensional Euclidean space in Eq. (2.1),
and evaluating some numerical factors, one obtains ($\epsilon_g =
V(D)$)

\begin{equation}
\epsilon_g = - {1 \over \pi^2} \int dq^2 \ q^2 \left[ \ln [1 + 3
d(q^2, \xi)] - {3 \over 4}d(q^2, \xi) + a \right],
\end{equation}
where constant $a = (3/4) - 2 \ln 2 = - 0.6363$ and the
integration from zero to infinity is assumed. The VED $\epsilon_g$
derived in Eq. (2.4) is already colorless quantity, since it has
been already summed over color indices. It does not also depend
explicitly on the unphysical (longitudinal) part of the full gluon
propagator due to the product $(D_0^{-1}D)$, which in its turn
comes from the above-mentioned normalization to zero (see above).

However, it still suffers from the two serious problems. The
coefficient of the transversal Lorentz structure (the effective
charge $d(q^2, \xi)$) may still depend explicitly on $\xi$. Also,
it is badly divergent at least as the fourth power of the
ultraviolet (UV) cutoff, and therefore suffers from different
types of the PT contributions (see below).

\section{The truly NP VED}

In order to define the VED free of all the above-mentioned
problems, let us make first the identical transformation of the
full gluon form factor (which is nothing but the effective charge)
in Eq. (2.2) as follows (Euclidean signature already):

\begin{equation}
d(q^2, \xi) = d(q^2, \xi) - d^{PT}(q^2, \xi)  + d^{PT}(q^2, \xi) =
d^{NP}(q^2, \xi)  + d^{PT}(q^2, \xi),
\end{equation}
where $d^{PT}(q^2, \xi)$ correctly describes the PT structure of
the effective charge $d(q^2, \xi)$, including its behavior in the
UV limit (asymptotic freedom \cite{12}), otherwise remaining
arbitrary. On the other hand, $d^{NP}(q^2, \xi)$, defined by the
above-made subtraction, is assumed to correctly reproduce the NP
structure of the effective charge, including its asymptotic in the
deep infrared (IR) limit, underlying thus the strong intrinsic
influence of the IR properties of the theory on its NP dynamics.
Evidently, both terms are valid in the whole energy/momentum
range, i.e, they are not asymptotics. Let us also emphasize the
principle difference between $d(q^2, \xi)$ and $d^{NP}(q^2, \xi)$.
The former is the NP quantity "contaminated" by the PT
contributions, while the latter one being also NP, nevertheless,
is free of them. Thus the exact separation between the truly NP
effective charge $d^{NP}(q^2, \xi)$ and its PT counterpart
$d^{PT}(q^2, \xi)$ is achieved (for the concrete example
considered below this separation is unique as well).

There is also another serious reason for the subtraction (3.1).
The problem is that the UV asymptotic of the full effective charge
due to asymptotic freedom may depend on the gauge-fixing
parameter, namely to leading order

\begin{equation}
d(q^2, \xi) \sim_{q^2 \rightarrow \infty} (\ln (q^2 /
\Lambda^2_{QCD}))^{c_0/b_0},
\end{equation}
where the exponent $c_0/b_0$ explicitly depends on the gauge-
fixing parameter via the coefficient $c_0$ \cite{12}, and
$\Lambda^2_{QCD}$ is the QCD asymptotic scale parameter.
Evidently, in the decomposition (3.1) precisely the PT part of the
full effective charge will be responsible for this explicit
dependence on the gauge choice. Subtracting it, we will be
guaranteed that the remaining part will not explicitly depend on
the gauge-fixing parameter (see below).

Substituting the exact decomposition (3.1) into Eq. (2.4) and
doing some trivial rearrangement, one obtains

\begin{equation}
\epsilon_g = - {1 \over \pi^2} \int dq^2 \ q^2 \left[ \ln [1 + 3
d^{NP}(q^2, \xi)] - {3 \over 4}d^{NP}(q^2, \xi) \right] +
\epsilon_{PT},
\end{equation}
where we introduce the following notation

\begin{equation}
\epsilon_{PT} = - {1 \over \pi^2}  \int dq^2 \ q^2 \left[ \ln [1 +
{3d^{PT}(q^2, \xi) \over 1 + 3 d^{NP}(q^2, \xi)}] - {3 \over
4}d^{PT}(q^2, \xi) + a \right].
\end{equation}
It contains the contribution which is mainly determined by the PT
part of the full gluon propagator, $d^{PT}(q^2, \xi)$. The
constant $a$ also should be included, since it comes from the
normalization of the free PT vacuum to zero. However, this is not
the whole story yet. The first term in Eq. (3.3), depending only
on the truly NP effective charge, nevertheless, assumes the
integration over the PT region (up to infinity), which should be
subtracted as well. The problem is that the first term in Eq.
(3.3) may still be divergent in the UV limit. $ A \ priori$, there
is only one restriction on its UV asymptotic, namely do not
contradict to asymptotically free behavior of the full effective
charge, to leading order shown in Eq. (3.2). For example, if
$d^{NP}(q^2, \xi)$ is a linearly falling at infinity function,
then the integral in Eq. (3.3) still diverges as mentioned above.

If we separate the NP region from the PT one, by introducing the
so-called effective scale $q_{eff}^2$ explicitly, then we get

\begin{equation}
\epsilon_g = - {1 \over \pi^2} \int_0^{q^2_{eff}} dq^2 \ q^2
\left[ \ln [1 + 3 d^{NP}(q^2, \xi)] - {3 \over 4}d^{NP}(q^2, \xi)
\right] + \epsilon_{PT} + \epsilon'_{PT},
\end{equation}
where evidently

\begin{equation}
\epsilon'_{PT} = - {1 \over \pi^2} \int_{q^2_{eff}}^{\infty} dq^2
\ q^2 \left[ \ln [1 + 3 d^{NP}(q^2, \xi)] - {3 \over 4}d^{NP}(q^2,
\xi) \right].
\end{equation}
This integral represents contribution to the VED which is
determined by the truly NP part of the full gluon propagator but
integrated out over the PT region. Along with $\epsilon_{PT}$
given in Eq. (3.4) it also represents a type of the PT
contribution into the gluon part of the VED (3.5). This means that
the two remaining terms in Eq. (3.5) should be subtracted by
introducing the truly NP YM VED $\epsilon_{YM}$ as follows:

\begin{equation}
\epsilon_{YM} = \epsilon_g - \epsilon_{PT} - \epsilon'_{PT},
\end{equation}
where the explicit expression for $\epsilon_{YM}$ is given by the
integral in Eq. (3.5) (see below as well). Concluding this
section, let us only note that the above-mentioned necessary
subtractions can be done in a more sophisticated way by means of
ghost degrees of freedom \cite{9}. In other words, the ghost
degrees of freedom are to be included into the PT parts of the
VED. Evidently, the subtracted terms are of no importance for our
present consideration.

\section{The Bag constant}

The Bag constant (the so-called the Bag pressure) is defined as
the difference between the PT and the NP VED \cite{13}. So in our
notations for YM fields, and as it follows from the definition
(3.7), it is nothing but the truly NP YM VED apart from the sign,
i.e.,

\begin{eqnarray}
B_{YM} = - \epsilon_{YM} &=& \epsilon_{PT} + \epsilon'_{PT} -
\epsilon_g \nonumber\\
&=& {1 \over \pi^2} \int_0^{q^2_{eff}} dq^2 \ q^2 \left[ \ln [1 +
3 d^{NP}(q^2, \xi)] - {3 \over 4}d^{NP}(q^2, \xi) \right].
\end{eqnarray}
This is a general expression for any model effective charge in
order to calculate the Bag constant (or the truly NP YM VED apart
from the sign) from first principles. It is our definition of the
truly NP YM VED and thus of the Bag constant (apart from the sign)
as integrated out the truly NP effective charge over the NP region
(soft momentum region, $0 \leq q^2 \leq q^2_{eff}$). It is free of
all types of the PT contributions, by construction. In this
connection, let us recall that $\epsilon_g$ is also NP, but
"contaminated" by the PT contributions, which just to be
subtracted in order to get expression (4.1).

Comparing expressions (2.4) and (4.1), one comes to the following
$prescription$ to derive Eq. (4.1) directly from Eq. (2.4).

(i). Replacing $d(q^2) \rightarrow d^{NP}(q^2)$.

(ii). Omitting the constant $a$ which normalizes the free PT
vacuum to zero.

(iii). Introducing the effective scale $q^2_{eff}$ which separates
the NP region from the PT one.

(iv). Omitting the minus sign for the Bag constant.

At this stage the Bag constant (4.1) is free of all types of the
PT "contaminations", and it is finite quantity, for sure. All
other its properties mentioned above (positivity, no imaginary
part, etc.) depend on the chosen effective charge. It is worth
emphasizing once more that the Bag constant and hence the truly NP
VED is the main physical characteristic of the true QCD ground
state. Let us note that in defining correctly the truly NP YM VED
(or equivalently the Bag constant), the three types of the
corresponding subtractions have been introduced. The first one -
in Eq. (3.1) at the fundamental gluon level and the two others -
in Eq. (3.7). For the general discussion of necessity in such
types of the subtractions see Ref. \cite{14}.

The quantum part of the total truly NP VED at log-loop level is

\begin{equation}
\epsilon_t = \epsilon_{YM} + N_f \epsilon_q,
\end{equation}
where $\epsilon_q$ is the truly NP quark loop contribution. It is
an order of magnitude less than $\epsilon_{YM}$ because of much
less quark degrees of freedom in the vacuum, and it is positive
because of overall minus due to quark loop. Evidently, in terms of
the Bag constant, one obtains

\begin{equation}
\epsilon_t = - B_{YM}[1 - N_f \nu],
\end{equation}
where we introduce $\epsilon_q = \nu B_{YM}$ and $\nu \ll 1$.

\section{Numerical evaluation of the Bag constant}

Eq. (4.1) is the main subject of our consideration in what
follows. The only problem remaining to solve is to choose such
truly NP effective charge $d^{NP}(q^2, \xi)$ which should not
$explicitly$ depend on the gauge-fixing parameter $\xi$. Here it
is worth emphasizing that the implicit gauge dependence is not
problem. Such kind of the dependence is unavoidable in gauge
theories like QCD, since fields themselves are gauge-dependent
\cite{15}. For different truly NP effective charges $d^{NP}(q^2,
\xi)$ we will get different numerical results. That is why the
choice for its explicit expression should be physically and
mathematically well justified.

For concrete numerical evaluation of the Bag constant (4.1), let
us choose the truly NP effective charge as follows:

\begin{equation}
d^{NP}(q^2) \equiv \alpha^{NP}_s(q^2) = \Lambda^2_{NP}/ q^2,
\end{equation}
where $\Lambda_{NP}$ is the mass scale parameter (the mass gap)
responsible for the large-scale structure of the true QCD vacuum.
First of all, in continuous QCD it is the NP solution of the
Schwinger-Dyson (SD) equation for the full gluon propagator
\cite{14} (and references therein). It is well known that in this
theory it leads to the linear rising potential between heavy
quarks, "seen" by lattice QCD \cite{16} as well ($(q^2)^{-2}$-type
behavior for the full gluon propagator). Secondly, it does not
depend explicitly on the gauge choice. Also, the separation
between the truly NP and PT effective charge is not only exact but
is unique as mentioned above, since the PT effective charge is
always regular at zero, while the truly NP effective charge (5.1)
is singular at origin. Let us note in advance that it leads to
many other desirable properties for the Bag constant (see below).

In order to proceed to the explicit numerical calculation of the
Bag constant for the truly NP effective charge (5.1), let us
introduce now dimensionless variable and parameter as follows:

\begin{equation}
z=q^2 / \Lambda^2_{NP}, \quad z_c=q_{eff}^2 / \Lambda^2_{NP},
\quad d^{NP}(q^2)=d^{NP}(z) = 1 / z.
\end{equation}
From the general expression for the Bag constant (4.1) one then
gets the dimensionless truly NP YM effective potential defined at
a fixed effective scale squared $q^4_{eff}$ as follows:

\begin{equation}
\Omega_{YM} (z_c) = { 1 \over q_{eff}^4} \epsilon_{YM}(q^4_{eff},
z_c) = {1 \over \pi^2} z_c^{-2}  \int_0^{z_c} dz \ z \left[ {3
\over 4}d^{NP}(z) - \ln [1 + 3 d^{NP}(z)] \right],
\end{equation}
which obviously makes it possible to factorize the scale
dependence in the truly NP YM VED (4.1). Performing almost trivial
integration in this expression, one obtains

\begin{equation}
\Omega_{YM} (z_c) = {1 \over 2 \pi^2} z_c^{-2} \left[ 9 \ln
\left(1 + {z_c \over 3 } \right)  - {3 \over 2} z_c - z_c^2 \ln
\left(1 + {3 \over z_c} \right) \right].
\end{equation}
It is easy to see now that as a function of $z_c$, the effective
potential (5.4) diverges as $\sim z_c^{-1}$ at small $z_c$ (the
unphysical regime, $\Lambda_{NP} \rightarrow \infty$ because the
mass gap is either finite or zero, i.e., it cannot be infinitely
large). It converges as $\sim - z_c^{-1}$ at infinity (PT limit,
$\Lambda_{NP} \rightarrow 0$), i.e. it approaches zero from below.
Let us emphasize that in the PT limit the truly NP phase vanishes
and the PT phase survives only. At a fixed effective scale
$q^2_{eff}$ it follows that $z_c \rightarrow \infty$, indeed. In
other words, at a fixed effective scale one recovers correct PT
limit, while at a fixed mass gap it evidently cannot be recovered.

From the above one can conclude that the effective potential (5.4)
as a function of $z_c$ has a minimum at some finite point (local
minimum), so its minimization makes sense. In general, by taking
first derivatives of the effective potential one recovers the
corresponding equations of motion \cite{10,11}. Thus the
above-mentioned minimization of the effective potential (5.4)
makes it possible to fix the constant of integration which is
$z_c$ in our case. Requiring $\partial \Omega_{YM} (z_c) /
\partial z_c = 0$, yields the following "stationary" condition

\begin{equation}
z_c = 4 \ln [1 + (z_c / 3)].
\end{equation}
Its numerical solution is

\begin{equation}
z_c^{min} = 2.2,
\end{equation}
so at the "stationary" state the effective potential (5.3) can be
written down as follows:

\begin{equation}
\Omega_{YM} (z_c^{min}) = { 1 \over 2 \pi^2} \left[ {3 \over 4}
(z_c^{min})^{-1} -  \ln \left(1 + {3 \over z_c^{min}} \right)
\right] = - 0.0263.
\end{equation}
The truly NP YM VED (5.3) and hence the Bag constant thus becomes

\begin{equation}
B_{YM} = - \epsilon_{YM} = 0.0263 q_{eff}^4 = 0.1273 \times
\Lambda^4_{NP},
\end{equation}
where the relation

\begin{equation}
q_{eff}^2 = z_c^{min} \Lambda^2_{NP} = 2.2 \Lambda^2_{NP}
\end{equation}
has been already used.

So, we have explicitly demonstrated that the Bag constant (5.8) at
the "stationary" state is finite, positive, and it has no
imaginary part, indeed. It depends only on the mass gap
responsible for the truly NP dynamics in the QCD ground state, or
equivalently on the effective scale separating the NP region from
the PT one. Evidently, only the minimization of the effective
potential (5.4) makes it possible to fix exactly the numerical
relation (5.9) between them.

\section{Numerical results}

In order to complete the numerical calculation of the Bag constant
(5.8) all we need now is the concrete value for the effective
scale $q_{eff}$, which separates the NP region from the PT one.
Equivalently, the concrete value for a a scale at which the NP
effects become important, that is the mass gap $\Lambda_{NP}$,
also allows one to achieve the same goal.

If the PT region starts conventionally from $1 \ GeV$, then this
number is a natural choice for the effective scale. It makes it
also possible to directly compare our values with the values of
many phenomenological parameters calculated just at this scale
(see Appendix A below). We consider this value as well justified
and realistic upper limit for the effective scale defined above.
Thus, we put

\begin{equation}
q_{eff} = q_{eff}^2 = 1 \ GeV,
\end{equation}
then for the mass gap, on account of the relation  (5.8), we get

\begin{equation}
\Lambda_{NP} = 0.6756 \ GeV.
\end{equation}

Equivalently, the numerical value of the mass gap $\Lambda_{NP}$
has been obtained from the experimental value for the pion decay
constant, $F_{\pi} =93.3 \ MeV$, by implementing a physically
well-motivated scale-setting scheme \cite{17,18}. The pion decay
constant is good experimental number, since it is directly
measured quantity in comparison, for example with the quark or the
gluon condensates, etc. For the mass gap we have obtained the
following numerical result:

\begin{equation}
\Lambda_{NP} = 0.5784 \ GeV,
\end{equation}
which for the effective scale yields

\begin{equation}
q_{eff} = 0.857 \ GeV, \quad q_{eff}^2 = 0.735 \ GeV^2.
\end{equation}
In what follows we will consider this value as a realistic lower
limit for the effective scale. One has to conclude that we have
obtained rather close numerical results for the effective scale
and the mass gap, by implementing rather different scale-setting
schemes. It is worth emphasizing that the effective scale (6.4)
quite well covers not only the deep IR region but the substantial
part of the intermediate one as well.

For the Bag constant (and hence for the truly NP YM VED) from Eq.
(5.8), one obtains

\begin{eqnarray}
B_{YM} = - \epsilon_{YM} &=& (0.0142-0.0263) \ GeV^4 \nonumber\\
&=& (1.84-3.4) \ GeV/fm^3 = (1.84-3.4) \times 10^{39} \ GeV/cm^3.
\end{eqnarray}
This is a huge amount of energy stored in one $cm^3$ of the NP QCD
vacuum even in "God-given" units $\hslash =c=1$. In order to
restore the explicit dependence on $\hbar$ \cite{10}, the
right-hand-side of Eq. (6.5) should be multiplied by it, which
numerical value in different units is $\hbar = 1 \times 10^{-34} J
\ s  = 6.6 \times 10^{-25} GeV \ s$. Evidently, using the number
of different conversion factors (see, for example Ref. \cite{15}
or Particle Data Group) the Bag constant can be expressed in
different systems of units (SI, CGS, etc.).

Taking into account that

\begin{equation}
1 \ GeV = 0.15 \times 10^{-9} J = 0.48 \times 10^{-17} \ W,
\end{equation}
from Eq. (6.5) one finally gets ($1 \ W = 10^{-3} kW = 10^{-6} MW
= 10^{-9} GW$)

\begin{equation}
B_{YM} = (0.88-1.63) \times 10^{13} \ GW /cm^3,
\end{equation}
in familiar units of watt (W). This number still indicates a huge
amount of energy stored in one $cm^3$ of the NP QCD vacuum
especially in comparison with the total nuclear energy produced in
one year and which is estimated about $360 \ GW$ only.

\section{Discussion}

We have already pointed out that the VED is badly divergent as the
fourth power of the UV cutoff, so we can write

\begin{equation}
\epsilon_g = - \Lambda^4_{UV} = - \lambda^4 B,
\end{equation}
which can be always done without loosing generality, and where
$\lambda$ is the dimensionless UV cutoff. At the same time, we
have obtained that the truly NP VED is finite and it is
$\epsilon_{YM} = - B$. This means that the further analysis can be
done in terms of the Bag constant.

Let us imagine now that we can extract the finite portion
$\epsilon_{YM}$ from the vacuum in $k$ different places (different
"vacuum energy releasing facilities" (VERF)). It can be done by
$n_m$ times in each place, where $m=1,2,3...k$. Then the remaining
(R) VED becomes

\begin{equation}
\epsilon_R =\epsilon_g - \epsilon_E = \epsilon_g - \sum_{m=1}^k
n_m \epsilon_{YM} =\epsilon_g + B \sum_{m=1}^k n_m,
\end{equation}
where, evidently, $\epsilon_E$ denotes the extracting (E) VED.

 The ideal case (which, however, will never be achieved) is
when we could extract the finite portion of the VED an infinitely
number times and in an infinitely number places, so we should
consider the following limits:

\begin{equation}
\epsilon_R = B \lim_{(\lambda, k, n_m) \rightarrow \infty} \Bigl[
- \lambda^4 + \sum_{m=1}^k n_m \Bigr].
\end{equation}
Thus, one can conclude that in the ideal case

\begin{equation}
\epsilon_E = B \lim_{(\lambda, k, n_m) \rightarrow \infty}
\sum_{m=1}^k n_m  \sim B \times \lambda^2,
\end{equation}
diverges quadratically since due to the limits pointed out in Eq.
(7.3), we can approximate $k \sim \lambda, \ n_m \sim \lambda, \
\lambda \rightarrow \infty$. At the same time, the initial VED
$\epsilon_g$ diverges quartically, i.e., $\epsilon_g \sim - B
\times \lambda^4, \ \lambda \rightarrow \infty$ (see Eq. (7.1)).
This means that the remaining VED

\begin{equation}
\epsilon_R = \epsilon_g - \epsilon_E \sim B \lim_{\lambda
\rightarrow \infty} [ - \lambda^4 + \lambda^2],
\end{equation}
will be always equal to the initial VED $\epsilon_g$ in the limit
pointed out above, i.e., in units of $B$ one has

\begin{equation}
\epsilon_R = \epsilon_g - \epsilon_E =  \epsilon_g + O(1/
\lambda^2), \quad \lambda \rightarrow \infty.
\end{equation}
This will be especially so in the real case when the integer
numbers $k$ and $n_m$ are very big, but finite. The true QCD
vacuum is an infinite and permanent source of energy. The only
problem is how to extract it in the profitable way. This is, of
course, beyond the scope of this paper, though there are some
preliminary ideas on this and related issues.

It is instructive to show explicitly that the same relation (7.6)
takes place between not only the different VED's but between
different types of energy themselves. Indeed, the different VED's
should be multiplied by volume (V) in order to get energy. Then
from Eqs. (6.5) and (7.4) in units of $[(1.84-3.4) \ GeV]$ it
follows that

\begin{equation}
E_E = V \epsilon_E = -(1.84-3.4) \ GeV  {V \over fm^3 } \times
\lim_{(\lambda, k, n_m) \rightarrow \infty} \sum_{m=1}^k n_m \sim
- \lambda^5,
\end{equation}
since $V / fm^3 \sim \lambda^3$ always when $\lambda$ goes to
infinity.

In the same units, one gets

\begin{equation}
E_g = V \epsilon_g = -(1.84-3.4) \ GeV  {V \over fm^3 } \times
\lambda^4 \sim - \lambda^7.
\end{equation}
So the difference between them which is nothing but the remaining
energy $E_R$ becomes

\begin{equation}
E_R = E_g - E_E =  E_g + O(1/ \lambda^2), \quad \lambda
\rightarrow \infty,
\end{equation}
in complete agreement with the relation (7.6) as it should be. Let
us remind that $[(1.84-3.4) \ GeV]$ units are equal to $[B \
fm^3]$ units, so in fact this simple analysis is again performed
in terms of the Bag constant.

\section{Conclusions}

In summary, we have formulated a general method as to how to
numerically calculate the quantum part of the truly NP YM VED (the
Bag constant, apart from the sign, by definition) in the covariant
gauge QCD ground state, using the effective potential approach for
composite operators. The Bag constant is defined as integrated out
the truly NP part of the effective charge over the NP region (soft
momentum region), Eq. (4.1). At this general stage the Bag
constant is colorless (color-singlet), finite, and it is free of
all types of the PT contributions, by construction. The separation
of "truly NP versus PT" contributions into the VED is exact and
unique because of the subtraction at the fundamental level (3.1),
as well as due to all other subtractions explicitly shown in Eq.
(3.7).

For the considered truly NP effective charge (5.1) in addition it
is negative, and it has no imaginary part (stable vacuum). It is
also a manifestly gauge-invariant quantity (i.e., does not
explicitly depend on the gauge-fixing parameter as it is
required). The separation of "soft versus hard" gluon momenta is
also exact because of the minimization procedure. It becomes
possible since the effective potential (5.4) as a function of the
dimensionless effective scale $z_c$, which separates the NP region
form the PT one, has a local minimum. Our numerical value for the
Bag constant is in a good agreement with other NP quantities such
as the gluon condensate. In order to demonstrate this we have used
the trace anomaly relation, not applying to the weak coupling
solution for the $\beta$ function (see Appendix A).

Our method can be generalized on the multi-loop contributions in
the effective potential, as well as to take into account the quark
degrees of freedom. These terms, however, will produce numerically
the contributions an order of magnitude less, at least, in
comparison with the leading log-loop level gluon term. What is
necessary indeed, is to be able to extract the finite part of the
truly NP VED in a self-consistent and manifestly gauge-invariant
ways. Just this is provided by our method which thus can be
applied to any QCD vacuum quantum and classical models. It may
serve as a test of them, providing an exact criterion for the
separation "stable versus unstable" vacua. Using our method we
have already shown that the vacuum of classical dual Abelian Higgs
model with string and without string contributions is unstable
against quantum corrections \cite{19,20}.

The Bag constant calculated here is a manifestly gauge-invariant
and colorless finite quantity, i.e., it is a physical quantity.
That is why it makes sense to discuss its releasing from the
vacuum. That's the VED in general is badly divergent is not a
mathematical problem. This reflects an universal reality. Vacuum
is everywhere and it always exists. Our Universe in general and
our real word in particular is only its special type of excitation
due to the Big Bang. As underlined above, the vacuum is infinite
and hence permanent source of energy. The only problem is how to
release the finite portion (the Bag constant) and whether it will
be profitable or not by introducing some type of cyclic process.
"Perpetuum mobile" does not exist, but "perpetuum source" of
energy does exist and it is the QCD ground state. Analogously to
the Plank constant $\hbar$ which determines a quant of energy
emitted by quantum physical systems, the Bag constant determines a
quant of energy which can be released from the QCD vacuum.

A financial support from HAS-JINR Scientific Collaboration Fund
and Hungarian OTKA-T043455 (P. Levai) is to be acknowledged.

\appendix
\section{The trace anomaly relation}

The truly NP VED (and hence the Bag constant) is important in its
own right as the main characteristic of the QCD ground state.
Furthermore it assists in calculating such an important
phenomenological parameter as the gluon condensate, introduced in
the QCD sum rules approach to resonance physics \cite{21}. The
famous trace anomaly relation \cite{22} in the general case
(nonzero current quark masses $m_f^0$) is

\begin{equation}
\Theta_{\mu\mu} = {\beta(\alpha_s) \over 4 \alpha_s} G^a_{\mu\nu}
G^a_{\mu\nu} + \sum_f m_f^0 \overline q_f q_f.
\end{equation}
where $\Theta_{\mu\mu}$ is the trace of the energy-momentum tensor
and $G^a_{\mu\nu}$ being the gluon field strength tensor while
$\alpha_s = g^2/4 \pi$. Sandwiching Eq. (A.1) between vacuum
states and on account of the obvious relation $\langle{0} |
\Theta_{\mu\mu} | {0}\rangle = 4 \epsilon_t$, one obtains

\begin{equation}
4 \epsilon_t =  \langle{0} | {\beta (\alpha_s)  \over 4 \alpha_s}
G^a_{\mu\nu} G^a_{\mu\nu} | {0}\rangle +  \sum_f m^0_f \langle{0}
| \overline q_f q_f | {0}\rangle.
\end{equation}
Here $\epsilon_t$ is the sum of all possible independent, NP
contributions to the VED (the total VED) and $\langle{0} |
\overline q_f q_f | {0}\rangle$ is the chiral quark condensate.
From this equation in the case of the pure YM fields (i.e., when
the number of quark fields is zero $N_f=0$), one obtains

\begin{equation}
\langle{0} | {\beta(\alpha_s) \over 4 \alpha_s} G^a_{\mu\nu}
G^a_{\mu\nu} | {0}\rangle = 4 \epsilon_{YM},
\end{equation}
where, evidently, we saturate the total VED $\epsilon_t$ by the
truly NP YM VED $\epsilon _{YM}$ defined in Eq. (4.1) and
calculated in Eq. (6.5) at log-loop level, i.e., setting
$\epsilon_t = \epsilon_{YM} + ...$. If confinement happens then
the $\beta$ function is always in the domain of attraction (i.e.,
always negative) without IR stable fixed point \cite{12}.
Therefore it is convenient to introduce the general definition for
the gluon condensate (i.e., without using the weak coupling limit
solution to the $\beta$ function) as follows:

\begin{equation}
\langle G^2 \rangle \equiv  - \langle{0} | {\beta(\alpha_s) \over
4 \alpha_s} G^a_{\mu\nu} G^a_{\mu\nu} | {0}\rangle =  - 4
\epsilon_{YM} = 4B_{YM}.
\end{equation}
Such defined general gluon condensate will be always positive as
it should be.

The renormalization group equation for the $\beta$ function

\begin{equation}
q^2 {d \alpha_s(q^2) \over d q^2} = \beta(\alpha_s(q^2))
\end{equation}
after substitution of our solution for the truly NP effective
charge (5.1) yields

\begin{equation}
\beta(\alpha_s(q^2)) = - \alpha_s(q^2),
\end{equation}
so that $\beta(\alpha_s) / \alpha_s = \beta(\alpha_s(q^2)) /
\alpha_s(q^2) =-1$, and Eq. (A.4) numerically becomes

\begin{equation}
\langle G^2 \rangle \equiv  - \langle{0} | {\beta(\alpha_s) \over
4 \alpha_s} G^a_{\mu\nu} G^a_{\mu\nu} | {0}\rangle = \langle{0} |
{1 \over 4 } G^a_{\mu\nu} G^a_{\mu\nu} | {0}\rangle = 4B_{YM} =
(0.0568-0.1032) \ GeV^4,
\end{equation}
where we have used the numerical value for the Bag constant (6.5).
To the gluon condensate (A.7) it can be assigned a physical
meaning indeed as the global (average) vacuum characteristic which
measures a density of the NP gluon fields configurations in the
true QCD vacuum.

However, it cannot be directly compared with the phenomenological
values for the standard gluon condensate estimated within
different approaches \cite{23}. The problem is that it is
necessary to remember that any value at the scale $\Lambda_{NP}$
(lower bound in the right-hand-side of Eq. (A.7)) is to be
recalculated at the $1 \ GeV$ scale. Moreover, both values
explicitly shown in Eq. (A.7) should be recalculated at the same
parametrization, which is nothing but the ratio $\beta(\alpha_s) /
\alpha_s$. In phenomenology the standard parametrization for the
gluon condensate is

\begin{equation}
G_2 = \langle {\alpha_s \over \pi} G^2 \rangle = \langle{0} |
{\alpha_s \over \pi} G^a_{\mu\nu} G^a_{\mu\nu} | {0}\rangle
\approx 0.012 \ GeV^4,
\end{equation}
which can be changed within a factor of 2 \cite{19}.

Thus in order to achieve the same parametrization the both sides
of Eq. (A.7) should be multiplied by $4(\alpha_s / \pi)$. For the
numerical value of the strong fine structure constant we use
$\alpha_s = \alpha_s(m_Z) = 0.1187$ from the Particle Data Group.
In addition the lower bound should be multiplied by the factor $ 1
\ CeV / \Lambda_{NP}=1.57$, coming form the numerical value (6.3).
Then the recalculated gluon condensate (A.7), which is denoted as
$\bar G_2$, finally becomes

\begin{equation}
\bar G_2 = (0.014-0.022) \ GeV^4.
\end{equation}
Just this numerical value for the gluon condensate should be
compared with the numerical value coming from the phenomenology,
see Eq. (A.8) above. So there is no doubt that all our numerical
results are in good agreement with other phenomenological
estimates \cite{23}.

\end{document}